\documentstyle[preprint,prb,aps]{revtex}
\begin{document}

\begin{titlepage}

\title
{Resonant quantum coherence of magnetization at
excited states in nanospin systems with different crystal symmetries}

\author{Jia-Lin Zhu, Rong L\"{u}\footnote {Author to whom 
the correspondence should be addressed.\\
Electronic address: rlu@castu.tsinghua.edu.cn}, and Su-Peng Kou} 
\address{Center for Advanced Study, 
Tsinghua University, Beijing 100084, People's Republic of China
}
\author{Hui Hu}
\address{
Department of Physics, 
Tsinghua University, Beijing 100084, People's Republic of China}
\author{Bing-Lin Gu}
\address{Center for Advanced Study, Tsinghua University,
Beijing 100084, People's Republic of China}

\maketitle
\begin{abstract}
The quantum interference effects induced by the Wess-Zumino term, or
Berry phase are studied theoretically in resonant quantum coherence
of the magnetization vector between degenerate states
in nanometer-scale single-domain ferromagnets in the absence of an external
magnetic field.
We consider the magnetocrystalline anisotropy
with trigonal, tetragonal and hexagonal crystal symmetry, respectively.
By applying the periodic instanton method in the
spin-coherent-state path integral, we
evaluate the low-lying tunnel splittings between degenerate
excited states of neighboring wells. And the low-lying
energy level spectrum of $m$-th excited state are obtained with the help
of the Bloch theorem in one-dimensional periodic potential. The
energy level spectrum and the thermodynamic properties of magnetic tunneling states
are found to depend significantly on the total spins of ferromagnets
at sufficiently low temperatures. Possible relevance to experiments is also
discussed.

\noindent
{\bf PACS number(s)}:  75.45.+j, 75.10.Jm, 03.65.Bz
\end{abstract}

\end{titlepage}

\section{Introduction}

In recent years there has been great experimental and theoretical effort to
observe and interpret macroscopic quantum tunneling (MQT) and coherence
(MQC) in nanometer-scale single-domain magnets.\cite{1} One notable subject
is that the topological Berry or Wess-Zumino phase\cite{2,3} can lead to
remarkable spin-parity effects. Loss {\it et al}.,\cite{4} and von Delft and
Henly\cite{5} showed that the tunnel splitting is suppressed to zero for
half-integer total spins in biaxial ferromagnetic (FM) particles due to the
destructive phase interfererence between topologically different tunneling
paths. However, the phase interference is constructive for integer spins,
and hence the splitting is nonzero.\cite{4,5} While spin-parity effects are
sometimes be related to Kramers degeneracy,\cite{4,5} they typically go
beyond the Kramers theorem in a rather unexpected way.\cite{6,7} Barnes {\it %
et al}. proposed the auxiliary particle method to study the model for a
single large spin subject to the external and anisotropy fields, and
discussed the spin-parity effects.\cite{8} Similar effect was found in
antiferromagnetic (AFM) particles, where only the integer excess spins can
tunnel but not the half-integer ones.\cite{11,12} Recently, topological
phase interference effects were investigated extensively in FM and AFM
particles in a magnetic field,\cite{6,9,10,13,14} and in the systems with
different symmetries.\cite{15,16,17} Spin tunneling and quantum oscillation
at excited states were studied for biaxial FM particles at zero magnetic
field,\cite{18} and at a field along the hard axis.\cite{19} One recent
experiment\cite{20} was performed to measure the tunnel splittings in
molecules Fe$_8$, and a clear oscillation of the splitting as a function of
the field along the hard axis was observed, which is a direct evidence of
the role of the topological spin phase (Berry phase) in the spin dynamics of
these molecules.

It is noted that the previous results of topological phase interference
effects were obtained for the tunnel splittings of the ground state in FM
particles with different crystal symmetries,\cite{16} or for the excited
states in FM particles with simple biaxial crystal symmetry.\cite{18} The
purpose of this paper is to study the spin-parity effects at excited states
for FM particles with a more complex (than biaxial) structure, such as
trigonal, tetragonal, and hexagonal symmetry around $\widehat{z}$, which
have three, four, and six degenerate easy directions in the basal plane.
Integrating out the momentum in the path integral, the spin tunneling
problem is mapped onto a particle moving problem in one-dimensional periodic
potential $V\left( \phi \right) $. By applying the periodic instanton
method, we obtain the low-lying tunnel splittings between $m$-th degenerate
excited states of neighboring wells. The periodic potential $V\left( \phi
\right) $ can be regarded as a one-dimensional superlattice. The general
translation symmetry results in the energy band structure, and the low-lying
energy level spectrum of excited states is obtained by using the Bloch
theorem and the tight-binding approximation. Our results show that the
excited-state tunnel splittings depend significantly on the parity of the
total spins. And the structure of energy level spectrum for the trigonal,
tetragonal and hexagonal crystal symmetry is found to be much more complex
than that for the biaxial crystal symmetry. Another important conclusion is
that the spin-parity effects can be reflected in thermodynamic quantities of
the low-lying tunneling levels. Thermodynamic property (such as the specific
heat) of the magnetic tunneling states is evaluated, and is found to be
strongly parity dependent on the total spins, which may provide an
experimental test for the topological phase interference effects. And the
spin-parity effect is lost at high temperatures.

The remaining part of this paper is organized as follows. In Sec. II, we
review briefly some basic ideas of MQT and MQC in FM\ particles, and discuss
the fundamentals concerning the computation of excited-level splittings in
the double-well-like potential. In Secs. III and IV, we study the spin
tunneling between degenerate excited states in FM particles with the
trigonal, tetragonal and hexagonal symmetry. The conclusions are presented
in Sec. V.

\section{Spin tunneling in FM particles}

For a spin tunneling problem, the tunnel splitting for MQC or the decay rate
for MQT is determined by the imaginary-time transition amplitude from an
initial state $\left| i\right\rangle $ at $\tau =-T/2$ to a final state $%
\left| f\right\rangle $ at $\tau =T/2$ in the spin-coherent-state
representation as\cite{3,31,32} 
\begin{equation}
U_{fi}=\left\langle f\right| e^{-HT}\left| i\right\rangle =\int D\Omega \exp
\left( -S_E\right) ,  \eqnum{1}
\end{equation}
where $D\Omega =\sin \theta d\theta d\phi $. The paths appearing in Eq. (1)
are fixed at the end points $\tau =\pm T/2$. For a system with equivalent
double wells, we let $\left| E\right\rangle _{+}$ and $\left| E\right\rangle
_{-}$ be eigenstates of the same energy $E$ in the right- and left-hand
wells, respectively. The small contribution due to quantum tunneling leads
to the effect of level splitting $\Delta E$, which removes the asymptotic
degeneracy. The corresponding eigenstates of the Hamiltonian separate into
odd and even states $\left| E\right\rangle _o$ and $\left| E\right\rangle _e$
which are superpositions of $\left| E\right\rangle _{+}$, $\left|
E\right\rangle _{-}$ such that $\left| E\right\rangle _o=\frac 1{\sqrt{2}}%
\left( \left| E\right\rangle _{+}-\left| E\right\rangle _{-}\right) $, and $%
\left| E\right\rangle _e=\frac 1{\sqrt{2}}\left( \left| E\right\rangle
_{+}+\left| E\right\rangle _{-}\right) $ with eigenvalues $E\pm \Delta E$,
respectively. In the limit that $T\rightarrow \infty $, one expects that the
amplitude for the transition from state $\left| E\right\rangle _{-}$ in the
left-hand well to the state $\left| E\right\rangle _{+}$ in the right-hand
well in the time interval $T$ as $_{+}\left\langle E\right| e^{-HT}\left|
E\right\rangle _{-}\rightarrow \exp \left( -ET\right) \sinh \left( \Delta
ET\right) $. Therefore the tunnel splitting $\Delta E$ is obtained if the
transition amplitude can be calculated. The decay rate $\Gamma $ from the
metastable state for MQT can be evaluated by a similar procedure. The
Euclidean action $S_E$ in Eq. (1) is\cite{3,31,32} 
\begin{equation}
S_E\left( \theta ,\phi \right) =\frac V\hbar \int d\tau \left[ i\frac{M_0}%
\gamma \left( \frac{d\phi }{d\tau }\right) -i\frac{M_0}\gamma \left( \frac{%
d\phi }{d\tau }\right) \cos \theta +E\left( \theta ,\phi \right) \right] , 
\eqnum{2}
\end{equation}
where $M_0=\left| \overrightarrow{M}\right| =\hbar \gamma S/V$, $V$ is the
volume of the particle, $\gamma $ is the gyromagnetic ratio, and $S$ is the
total spins. It is noted that the first two terms in Eq. (2) define the
Berry or Wess-Zumino term which has a simple topological interpretation. For
a closed path, this term equals $-iS$ times the area swept out on the unit
sphere between the path and the north pole. The first term in Eq. (2) is a
total imaginary-time derivative, which has no effect on the classical
equations of motion, but it is crucial for the spin-parity effects.\cite
{4,5,21}

In the semiclassical limit, the instanton's contribution to $\Gamma $ or $%
\Delta E$ (not including the topological Wess-Zumino phase) is given by\cite
{22} 
\begin{equation}
\Gamma \ (\text{or }\ \Delta E)=A\omega _p\left( \frac{S_{cl}}{2\pi }\right)
^{1/2}e^{-S_{cl}},  \eqnum{3}
\end{equation}
where $\omega _p$ is the oscillation frequency in the well, $S_{cl}$ is the
classical action, and the prefactor $A$ originates from the quantum
fluctuations about the classical path. It is noted that Eq. (3) is based on
tunneling at the ground state, and the temperature dependence of the
tunneling frequency (i.e., tunneling at excited states) is not taken into
account. The instanton technique is suitable only for the evaluation of the
tunneling rate at the vacuum level, since the usual (vacuum) instantons
satisfy the vacuum boundary conditions. Recently, Liang {\it et al}.\cite{23}
developed new types of pseudoparticle configurations which satisfy periodic
boundary condition (i.e., periodic instantons or nonvacuum instantons). For
a particle moving in a double-well-like potential $U\left( x\right) $, the
WKB method gives the tunnel splitting of excited states at an energy $E>0$ as%
\cite{24,25,33} 
\begin{equation}
\Delta E=\frac{\omega \left( E\right) }\pi \exp \left[ -S\left( E\right)
\right] ,  \eqnum{4}
\end{equation}
with the imaginary-time action is 
\begin{equation}
S\left( E\right) =2\sqrt{2m}\int_{x_1\left( E\right) }^{x_2\left( E\right)
}dx\sqrt{U\left( x\right) -E},  \eqnum{5}
\end{equation}
where $x_{1,2}\left( E\right) $ are the turning points for the particle
oscillating in the inverted potential $-U\left( x\right) $. $\omega \left(
E\right) =2\pi /t\left( E\right) $ is the energy-dependent frequency, and $%
t\left( E\right) $ is the period of the real-time oscillation in the
potential well, 
\begin{equation}
t\left( E\right) =\sqrt{2m}\int_{x_3\left( E\right) }^{x_4\left( E\right) }%
\frac{dx}{\sqrt{E-U\left( x\right) }},  \eqnum{6}
\end{equation}
where $x_{3,4}\left( E\right) $ are the classical turning points for the
particle oscillating inside $U\left( x\right) $. The functional-integral and
the WKB\ method showed that for the potentials parabolic near the bottom the
result Eq. (4) should be multiplied by $\sqrt{\frac \pi e}\frac{\left(
2n+1\right) ^{n+1/2}}{2^ne^nn!}$.\cite{25,33} This factor is very close to 1
for all $n$: 1.075 for $n=0$, 1.028 for $n=1$, 1.017 for $n=2$, etc.
Stirling's formula for $n!$ shows that this factor trends to 1 as $%
n\rightarrow \infty $. Therefore, this correction factor, however, does not
change much in front of the exponentially small action term in Eq. (4).
Recently, the crossover from quantum to classical behavior and the
associated phase transition have been investigated extensively in nanospin
systems\cite{25,26,27,28,29} and other systems.\cite{30}

\section{MQC for trigonal crystal symmetry}

In this section, we consider a spin system with trigonal crystal symmetry,
i.e., which has three consecutive energy minima in a period. Now the total
energy is 
\begin{equation}
E\left( \theta ,\phi \right) =K_1\cos ^2\theta -K_2\sin ^3\theta \cos \left(
3\phi \right) +E_0,  \eqnum{7}
\end{equation}
where $K_1$ and $K_2$ are the magnetic anisotropic constants satisfying $%
K_1\gg K_2>0$, and $E_0$ is a constant which makes $E\left( \theta ,\phi
\right) $ zero at the initial state. As $K_1\gg K_2>0$, the magnetization
vector is forced to lie in the $\theta =\pi /2$ plane, so the fluctuations
of $\theta $ about $\pi /2$ are small. Introducing $\theta =\pi /2+\alpha $ $%
\left( \left| \alpha \right| \ll 1\right) $, Eq. (7) reduces to 
\begin{equation}
E\left( \alpha ,\phi \right) \approx K_1\alpha ^2+2K_2\sin ^2\left( 3\phi
/2\right) .  \eqnum{8}
\end{equation}
The ground state corresponds to the magnetization vector pointing in one of
the three degenerate easy directions: $\theta =\pi /2$, and $\phi =0$, $2\pi
/3$, $4\pi /3$, other energy minima repeat the three states with period $%
2\pi $. Performing the Gaussian integration over $\alpha $, we can map the
spin system onto a particle moving problem in one-dimensional potential
well. Now the transition amplitude becomes 
\begin{eqnarray}
U_{fi} &=&\exp \left[ -iS\left( \phi _f-\phi _i\right) \right] \int d\phi
\exp \left( -S_E\left[ \phi \right] \right) ,  \nonumber \\
&=&\exp \left[ -iS\left( \phi _f-\phi _i\right) \right] \int d\phi \exp
\left\{ -\int d\tau \left[ \frac 12m\left( \frac{d\phi }{d\tau }\right)
^2+V\left( \phi \right) \right] \right\} ,  \eqnum{9}
\end{eqnarray}
with $m=\hbar S^2/2K_1V$, and $V\left( \phi \right) =2\left( K_2V/\hbar
\right) \sin ^2\left( 3\phi /2\right) $. It is noted that the total
derivative in Eq. (2), when integrated, gives an additional phase factor to
the transition amplitude Eq. (9) which depends on the initial and final
values of $\phi $. For the trigonal symmetry, this phase factor in Eq. (9)
is $\exp \left( -i2\pi S/3\right) $. The potential $V\left( \phi \right) $
is periodic with period $2\pi /3$, and there are three minima in the entire
region $2\pi $. We may look at $V\left( \phi \right) $ as a superlattice
with lattice constant $2\pi /3$ and total length $2\pi $, and we can derive
the energy spectrum by applying the Bloch theorem and the tight-binding
approximation. The translational symmetry is ensured by the possibility of
successive $2\pi $ extensions.

The periodic instanton configuration $\phi _p$ which minimizes the Euclidean
action in Eq. (9) satisfies the equation of motion 
\begin{equation}
\frac 12m\left( \frac{d\phi _p}{d\tau }\right) ^2-V\left( \phi _p\right) =-E,
\eqnum{10}
\end{equation}
where $E>0$ is a constant of integration, which can be viewed as the
classical energy of the pseudoparticle configuration. Then we obtain the
kink-solution as 
\begin{equation}
\sin ^2\left( \frac 32\phi _p\right) =1-k^2\text{sn}^2\left( \omega _1\tau
,k\right) ,  \eqnum{11}
\end{equation}
where sn$\left( \omega _1\tau ,k\right) $ is the Jacobian elliptic sine
function of modulus $k$, 
\begin{equation}
k^2=\frac{n_1^2-1}{n_1^2},  \eqnum{12}
\end{equation}
with $\omega _1=3\sqrt{2}\left( V/\hbar S\right) \sqrt{K_1K_2}$, and $n_1=%
\sqrt{2K_2V/\hbar E}>1$. In the low energy limit, i.e., $E\rightarrow 0$, $%
k\rightarrow 1$, sn$\left( u,1\right) \rightarrow \tanh u$, we have 
\begin{equation}
\sin ^2\left( \frac 32\phi _p\right) =\frac 1{\cosh ^2\left( \omega _1\tau
\right) },  \eqnum{13}
\end{equation}
which is exactly the vacuum instanton solution derived in Ref. 16.

The Euclidean action of the periodic instanton configuration Eq. (11) over
the domain $\left( -\beta ,\beta \right) $ is found to be 
\begin{equation}
S_p=\int_{-\beta }^\beta d\tau \left[ \frac 12m\left( \frac{d\phi _p}{d\tau }%
\right) ^2+V\left( \phi _p\right) \right] =W+2E\beta ,  \eqnum{14}
\end{equation}
with 
\begin{equation}
W=\frac{2^{5/2}}3S\sqrt{\frac{K_2}{K_1}}\left[ E\left( k\right) -\left(
1-k^2\right) K\left( k\right) \right] ,  \eqnum{15}
\end{equation}
where $K\left( k\right) $ and $E\left( k\right) $ are the complete elliptic
integral of the first and second kind, respectively. Now we discuss the low
energy limit where $E$ is much less than the barrier height. In this case, $%
k^{\prime 2}=1-k^2=\hbar E/2K_2V\ll 1$, so we can perform the expansions of $%
K\left( k\right) $ and $E\left( k\right) $ in Eq. (15) to include terms like 
$k^{\prime 2}$ and $k^{\prime 2}\ln \left( 4/k^{\prime }\right) $, 
\begin{eqnarray}
E\left( k\right) &=&1+\frac 12\left[ \ln \left( \frac 4{k^{\prime }}\right) -%
\frac 12\right] k^{\prime 2}+\cdots ,  \nonumber \\
K\left( k\right) &=&\ln \left( \frac 4{k^{\prime }}\right) +\frac 14\left[
\ln \left( \frac 4{k^{\prime }}\right) -1\right] k^{\prime 2}+\cdots . 
\eqnum{16}
\end{eqnarray}
With the help of small oscillator approximation for energy near the bottom
of the potential well, $E=\varepsilon _m^{tri}=\left( m+1/2\right) \omega _1$%
, Eq. (15) is expanded as 
\begin{equation}
W=\frac{2^{5/2}}3\sqrt{\frac{K_2}{K_1}}S-\left( m+\frac 12\right) +\left( m+%
\frac 12\right) \ln \left[ \frac 3{2^{9/2}}\sqrt{\frac{K_1}{K_2}}\frac 1S%
\left( m+\frac 12\right) \right] .  \eqnum{17}
\end{equation}
Then the general formula Eq. (4) gives the low-lying energy shift of $m$-th
excited states for FM particles with trigonal crystal symmetry at zero
magnetic field as 
\begin{equation}
\hbar \Delta \varepsilon _m^{tri}=\frac 1{m!}\frac{2^{\frac 92\left( m+\frac %
12\right) }}{3^{\left( m-\frac 12\right) }\times \pi ^{\frac 12}}\left(
K_1V\right) \lambda ^{\frac 12\left( m+\frac 32\right) }S^{\left( m-\frac 12%
\right) }\exp \left( -\frac{2^{5/2}}3\lambda ^{1/2}S\right) ,  \eqnum{18}
\end{equation}
where $\lambda =K_2/K_1$, and $S$ is the total spin.

It is noted that $\hbar \Delta \varepsilon _m^{tri}$ is only the level shift
induced by tunneling between degenerate excited states through a single
barrier. The periodic potential $V\left( \phi \right) $ can be regarded as a
one-dimensional superlattice. The general translation symmetry results in
the energy band structure, and the energy spectrum could be determined by
the Bloch theorem. It is easy to show that if $\varepsilon _m^{tri}$ are the
degenerate eigenvalues of the system with infinitely high barrier, the
energy level spectrum is given by the following formula with the help of
tight-binding approximation, 
\begin{equation}
E_m^{tri}=\varepsilon _m^{tri}-2\Delta \varepsilon _m^{tri}\cos \left[
\left( S+\xi \right) 2\pi /3\right] .  \eqnum{19}
\end{equation}
The Bloch wave vector $\xi $ can be assumed to take either of the three
values $-1$, $0$, $1$ in the first Brillouin zone. It is noted that in Eq.
(19) we have included the contribution of topological phase for FM\
particles with trigonal crystal symmetry (i.e., $2\pi S/3$).. The low-lying
energy level spectrum, which corresponds to the splittings of $m$-th excited
state due to the resonant quantum coherence of the magnetization vector
between energetically degenerate states, is found to depend on the parity of
total spin of FM particle significantly. If $S$ is an integer, the low-lying
energy level spectrum is $\hbar \varepsilon _m^{tri}-2\hbar \Delta
\varepsilon _m^{tri}$, and $\hbar \varepsilon _m^{tri}+\hbar \Delta
\varepsilon _m^{tri}$, the latter being doubly degenerate. But if $S$ is a
half-integer, the low-lying energy level spectrum is $\hbar \varepsilon
_m^{tri}-\hbar \Delta \varepsilon _m^{tri}$, and $\hbar \varepsilon
_m^{tri}+2\hbar \Delta \varepsilon _m^{tri}$, the former being doubly
degenerate. This spin-parity effect is the result of phase interference
between topologically distinct tunneling path.

At the end of this section, we discuss the possible relevance to the
experimental test for spin-parity effects in single-domain FM nanoparticles.
First we discuss the thermodynamic behavior of this system at very low
temperature $T\sim T_0=\hbar \Delta \varepsilon _0^{tri}/k_B$. For FM\
particles with trigonal crystal symmetry at such a low temperature, the
partition function of the ground state is found to be 
\begin{equation}
Z=\exp \left( -\beta \hbar \varepsilon _0^{tri}\right) \left[ \exp \left(
\pm 2\beta \hbar \Delta \varepsilon _0^{tri}\right) +2\exp \left( \mp \beta
\hbar \Delta \varepsilon _0^{tri}\right) \right] ,  \eqnum{20}
\end{equation}
where upper sign corresponds to integer spins, lower sign corresponds to
half-integer spins, and $\varepsilon _0^{tri}=\omega _1/2$. Then the
specific heat is $c=-T\left( \partial ^2F/\partial T^2\right) $, with $%
F=-k_BT\ln Z$. For the low temperature case, the result is 
\begin{equation}
c=18k_B\left( \beta \hbar \Delta \varepsilon _0^{tri}\right) ^2\frac{\exp
\left( \pm \beta \hbar \Delta \varepsilon _0^{tri}\right) }{\left[ \exp
\left( \pm 2\beta \hbar \Delta \varepsilon _0^{tri}\right) +2\exp \left( \mp
\beta \hbar \Delta \varepsilon _0^{tri}\right) \right] ^2}.  \eqnum{21}
\end{equation}
In Fig. 1, we plot the temperature dependence of the specific heat for
integer and half-integer spins at very low temperature $0\leq T/T_0\leq 1$.
It is clearly shown that the specific heat for integer spins is much
different from that for half-integer spins at sufficiently low temperatures.
When the temperature is higher $\hbar \Delta \varepsilon _0^{tri}\ll
k_BT<\hbar \omega _1$, the excited energy levels may give contribution to
the partition function. Now the partition function is 
\begin{equation}
Z\approx Z_0\left[ 1+\left( 1-e^{-\beta \hbar \omega _1}\right) \left( \beta
\hbar \Delta \varepsilon _0^{tri}\right) ^2I_0\left( 2q_1e^{-\beta \hbar
\omega _1/2}\right) \right] ,  \eqnum{22}
\end{equation}
for both integer and half-integer spins, where $Z_0=3e^{-\beta \hbar \omega
_1/2}/\left( 1-e^{-\beta \hbar \omega _1}\right) $ is the partition function
in the well calculated for $k_BT\ll \Delta U$ over the low-lying
oscillatorlike states with $\varepsilon _m^{tri}=\left( m+1/2\right) \omega
_1$, and $\omega _1=3\sqrt{2}\left( V/\hbar S\right) \sqrt{K_1K_2}$. $%
I_0\left( x\right) =\sum_{n=0}\left( x/2\right) ^{2n}/\left( n!\right) ^2$
is the modified Bassel function, and $q_1=\left( 2^{9/2}/3\right) \lambda
^{1/2}S>1$. We define a characteristic temperature $\widetilde{T}$ that is
solution of equation $q_1e^{-\hbar \omega _1/k_B\widetilde{T}}=1$. The
temperature $\widetilde{T}=\hbar \omega _1/2\ln q_1$ characterizes the
crossover from thermally assisted tunneling to the ground-state tunneling.
In Fig. 2, we plot the temperature dependence of the specific heat for
integer and half-integer spins at high temperature $30\leq T/T_0\leq 60$.
The result shows that the spin-parity effect will be lost at high
temperatures. The specific heat for integer spins is almost the same as that
for half-integer spins.

\section{MQC for tetragonal and hexagonal crystal symmetries}

In this section, we will apply the method in Sec. III to study spin
tunneling in FM particles with tetragonal and hexagonal crystal symmetry.
For the tetragonal symmetry, 
\begin{equation}
E\left( \theta ,\phi \right) =K_1\cos ^2\theta +K_2\sin ^4\theta
-K_2^{\prime }\sin ^4\theta \cos \left( 4\phi \right) +E_0,  \eqnum{23}
\end{equation}
where $K_1\gg K_2$, $K_2^{\prime }>0$. The energy minima of this system are
at $\theta =\pi /2$, and $\phi =0$, $\pi /2$, $\pi $, $3\pi /2$, and other
energy minima repeat the four states with period $2\pi $. The problem can be
mapped onto a problem of one-dimensional motion by integrating out the
fluctuations of $\theta $ about $\pi /2$, and for this case $V\left( \phi
\right) =2\left( K_2^{\prime }V/\hbar \right) \sin ^2\left( 2\phi \right) $.
Now $V\left( \phi \right) $ is periodic with period $\pi /2$, and there are
four minima in the entire region $2\pi $. The periodic instanton
configuration with an energy $E>0$ is $\sin ^2\left( 2\phi _p\right) =1-k^2$%
sn$^2\left( \omega _2\tau ,k\right) $, where $k=\sqrt{\left( n_1^2-1\right)
/n_1^2}$, $\omega _2=2^{5/2}\left( V/\hbar S\right) \sqrt{K_1K_2^{\prime }}$%
, and $n_1=\sqrt{2K_2^{\prime }V/\hbar E}>1$. The associated classical
action is $S_p=W+2E\beta $, with 
\begin{equation}
W=2^{1/2}S\sqrt{\frac{K_2^{\prime }}{K_1}}\left[ E\left( k\right) -\left(
1-k^2\right) K\left( k\right) \right] .  \eqnum{24}
\end{equation}
The general formula Eq. (4) gives the low-lying energy shift of $m$-th
excited state as 
\begin{equation}
\hbar \Delta \varepsilon _m^{te}=\frac 1{m!}\frac{2^{\frac 52m+\frac 94}}{%
\pi ^{\frac 12}}\left( K_1V\right) \lambda ^{\frac 12\left( m+\frac 32%
\right) }S^{\left( m-\frac 12\right) }\exp \left( -2^{1/2}\lambda
^{1/2}S\right) ,  \eqnum{25}
\end{equation}
with $\lambda =K_2^{\prime }/K_1$. The periodic potential $V\left( \phi
\right) $ can be viewed as a superlattice with lattice constant $\pi /2$ and
total length $2\pi $, and the Bloch theorem then gives the energy level
spectrum of $m$-th excited state $\varepsilon _m^{te}=\left( m+1/2\right)
\omega _2$ as $E_m^{te}=\varepsilon _m^{te}-2\Delta \varepsilon _m^{te}\cos
\left[ \left( S+\xi \right) \pi /2\right] $, where $\xi =-1,0,1,2$ in the
first Brillouin zone. It is easy to show that the low-lying energy level
spectrum is $\hbar \varepsilon _m^{te}\pm 2\hbar \Delta \varepsilon _m^{te}$%
, and $\hbar \varepsilon _m^{te}$ for integer spins, the latter being doubly
degenerate. While the level spectrum is $\hbar \varepsilon _m^{te}\pm \sqrt{2%
}\hbar \Delta \varepsilon _m^{te}$ with doubly degenerate for half-integer
spins. At a very low temperature $T\sim T_0=\hbar \Delta \varepsilon
_0^{te}/k_B$, the specific heat is 
\begin{equation}
c=4k_B\left( \beta \hbar \Delta \varepsilon _0^{te}\right) ^2\frac 1{1+\cosh
\left( 2\beta \hbar \Delta \varepsilon _0^{te}\right) },  \eqnum{26a}
\end{equation}
for integer spins, while 
\begin{equation}
c=2k_B\left( \beta \hbar \Delta \varepsilon _0^{te}\right) ^2\frac 1{\cosh
^2\left( \sqrt{2}\beta \hbar \Delta \varepsilon _0^{te}\right) }, 
\eqnum{26b}
\end{equation}
for half-integer spins. The tunneling behavior for integer spins is almost
the same as that for half-integer spins at high temperature $\hbar \Delta
\varepsilon _0^{te}\ll k_BT<\hbar \omega _2$.

For the case of hexagonal symmetry, 
\begin{equation}
E\left( \theta ,\phi \right) =K_1\cos ^2\theta +K_2\sin ^4\theta +K_3\sin
^6\theta -K_3^{\prime }\sin ^6\theta \cos \left( 6\phi \right) +E_0, 
\eqnum{27}
\end{equation}
where $K_1\gg K_2,K_3,K_3^{\prime }>0$. The easy directions are at $\theta
=\pi /2$, and $\phi =0$, $\pi /3$, $2\pi /3$, $\pi $, $4\pi /3$, $5\pi /3$,
and other energy minima repeat the six states with period $2\pi $. For the
present case, $V\left( \phi \right) =2\left( K_3^{\prime }V/\hbar \right)
\sin ^2\left( 3\phi \right) $ is periodic with period $\pi /3$, and there
are six minima in the entire region $2\pi $. The periodic instanton
configuration at a given energy $E>0$ is $\sin ^2\left( 3\phi _p\right)
=1-k^2$sn$^2\left( \omega _3\tau ,k\right) $, where $k=\sqrt{\left(
n_1^2-1\right) /n_1^2}$, $\omega _3=\left( 3\times 2^{3/2}\right) \left(
V/\hbar S\right) \sqrt{K_1K_3^{\prime }}$, and $n_1=\sqrt{2K_3^{\prime
}V/\hbar E}>1$. Correspondingly, the classical action is $S_p=W+2E\beta $,
with 
\begin{equation}
W=2^{3/2}S\sqrt{\frac{K_3^{\prime }}{K_1}}\left[ E\left( k\right) -\left(
1-k^2\right) K\left( k\right) \right] ,  \eqnum{28}
\end{equation}
and the low-lying energy shift of $m$-th excited state is 
\begin{equation}
\hbar \Delta \varepsilon _m^{he}=\frac 1{m!}\frac{2^{\frac 72m+\frac{11}4}}{%
3^{m-\frac 12}\pi ^{\frac 12}}\left( K_1V\right) \lambda ^{\frac 12\left( m+%
\frac 32\right) }S^{\left( m-\frac 12\right) }\exp \left( -\frac{2^{3/2}}3%
\lambda ^{1/2}S\right) ,  \eqnum{29}
\end{equation}
with $\lambda =K_3^{\prime }/K_1$. Now $V\left( \phi \right) $ can be
regarded as a one-dimensional superlattice with lattice constant $\pi /3$.
By applying the Bloch theorem and the tight-binding approximation, we obtain
the energy level spectrum of $m$-th excited state $\varepsilon
_m^{he}=\left( m+1/2\right) \omega _3$ as $E_m^{he}=\varepsilon
_m^{he}-2\Delta \varepsilon _m^{he}\cos \left[ \left( S+\xi \right) \pi
/3\right] $, where $\xi =-2,-1,0,1,2,3$. If $S$ is an integer, the low-lying
energy level spectrum is $\hbar \varepsilon _m^{he}\pm 2\hbar \Delta
\varepsilon _m^{he}$, and $\hbar \varepsilon _m^{he}\pm \hbar \Delta
\varepsilon _m^{he}$, the latter two levels being doubly degenerate. If $S$
is a half-integer, the level spectrum is $\hbar \varepsilon _m^{he}\pm \sqrt{%
3}\hbar \Delta \varepsilon _m^{he}$, and $\hbar \varepsilon _m^{he}$, all
three levels being doubly degenerate. Then the specific heat at sufficiently
low temperatures is 
\begin{equation}
c=2k_B\left( \beta \hbar \Delta \varepsilon _0^{he}\right) ^2\frac{\left[
4+4\cosh \left( \beta \hbar \Delta \varepsilon _0^{he}\right) +\cosh \left(
2\beta \hbar \Delta \varepsilon _0^{he}\right) \cosh \left( \beta \hbar
\Delta \varepsilon _0^{he}\right) \right] }{\left[ \cosh \left( 2\beta \hbar
\Delta \varepsilon _0^{he}\right) +2\cosh \left( \beta \hbar \Delta
\varepsilon _0^{he}\right) \right] ^2},  \eqnum{30a}
\end{equation}
for integer spins, while 
\begin{equation}
c=6k_B\left( \beta \hbar \Delta \varepsilon _0^{he}\right) ^2\frac{2+\cosh
\left( \sqrt{3}\beta \hbar \Delta \varepsilon _0^{he}\right) }{\left[
1+2\cosh \left( \sqrt{3}\beta \hbar \Delta \varepsilon _0^{he}\right)
\right] ^2},  \eqnum{30b}
\end{equation}
for half-integer spins.

In brief, the low-lying energy level spectrum and the heat capacity of the
magnetic tunneling states for tetragonal and hexagonal symmetry are found to
depend on the parity of the total spins, resulting from the Wess-Zumino
phase interference between topologically distinct tunneling paths. And this
spin-parity or topological phase interference effect will be lost at high
temperatures.

\section{Conclusions}

In summary, we have investigated the topological phase interference effects
in spin tunneling at excited levels for single-domain FM particles with
trigonal, tetragonal, and hexagonal crystal symmetries. The low-lying tunnel
splittings between $m$-th degenerate excited states of neighboring wells are
evaluated with the help of the periodic instanton method, and the energy
level spectrum is obtained by applying the Bloch theorem and the
tight-binding approximation in one-dimensional periodic potential.

One important conclusion is that for all the three kinds of crystal
symmetries, the low-lying energy level spectrum for integer total spins is
significantly different from that for half-integer total spins, resulting
from the Berry phase interference between topologically distinct tunneling
paths. For FM particles with simple biaxial crystal symmetry, which has two
degenerate easy directions in the basal plane (i.e., the double-well
system), it has been theoretically shown that the tunnel splitting is
suppressed to zero for half-integer spins due to the destructive phase
interference between topologically different tunneling paths connecting the
same initial and final states. However, the structure of low-lying tunneling
level spectrum for the trigonal, tetragonal, or hexagonal crystal symmetry
is found to be much more complex than that for the biaxial crystal symmetry.
The low-lying energy level spectrum can be nonzero even if the total spin is
a half-integer for the trigonal, tetragonal, or hexagonal crystal symmetry.
Note that these spin-parity effects are of topological origin, and therefore
are independent of the magnitude of total spins of FM particles. The heat
capacity of low-lying magnetic tunneling states is evaluated and is found to
depend significantly on the parity of total spins for FM particles with
different crystal symmetries at sufficiently low temperatures, providing a
possible experimental method to examine the theoretical results on
topological phase interference effects. And the spin-parity effects will be
lost at high temperatures. Our results presented here should be useful for a
quantitative understanding on the topological phase interference or
spin-parity effects in resonant quantum tunneling of magnetization in
single-domain FM particles with different crystal symmetries.

More recently, Wernsdorfer and Sessoli\cite{20} have measured the tunnel
splittings in the molecular Fe$_8$ clusters, and have found a clear
oscillation of the tunnel splitting with the field along hard axis, which is
a direct evidence of the role of the Berry phase in the spin dynamics of
these molecules. It is noted that the theoretical results presented in this
paper are based on the instanton method, which is semiclassical in nature,
i.e., valid for large spins and in the continuum limit. Whether the
instanton method can be applied in studying the spin dynamics in molecular
clusters with $S=10$ (such as Fe$_8$) is an open question.

The theoretical calculations performed in this paper can be extended to the
FM and AFM particles in a magnetic field. Work along this line is still in
progress. We hope that the theoretical results presented here will stimulate
more experiments whose aim is observing the topological phase interference
effects in nanometer-scale single-domain magnets.

\section*{Acknowledgments}

R. L. and S. P. Kou are grateful to Dr. Y. Zhou, Prof. Z. Xu, Prof. J. -Q.
Liang and Prof. F. -C. Pu for stimulating discussions. J. -L. Zhu and R. L.
would like to thank Prof. W. Wernsdorfer and Prof. R. Sessoli for providing
their paper (Ref. 20).

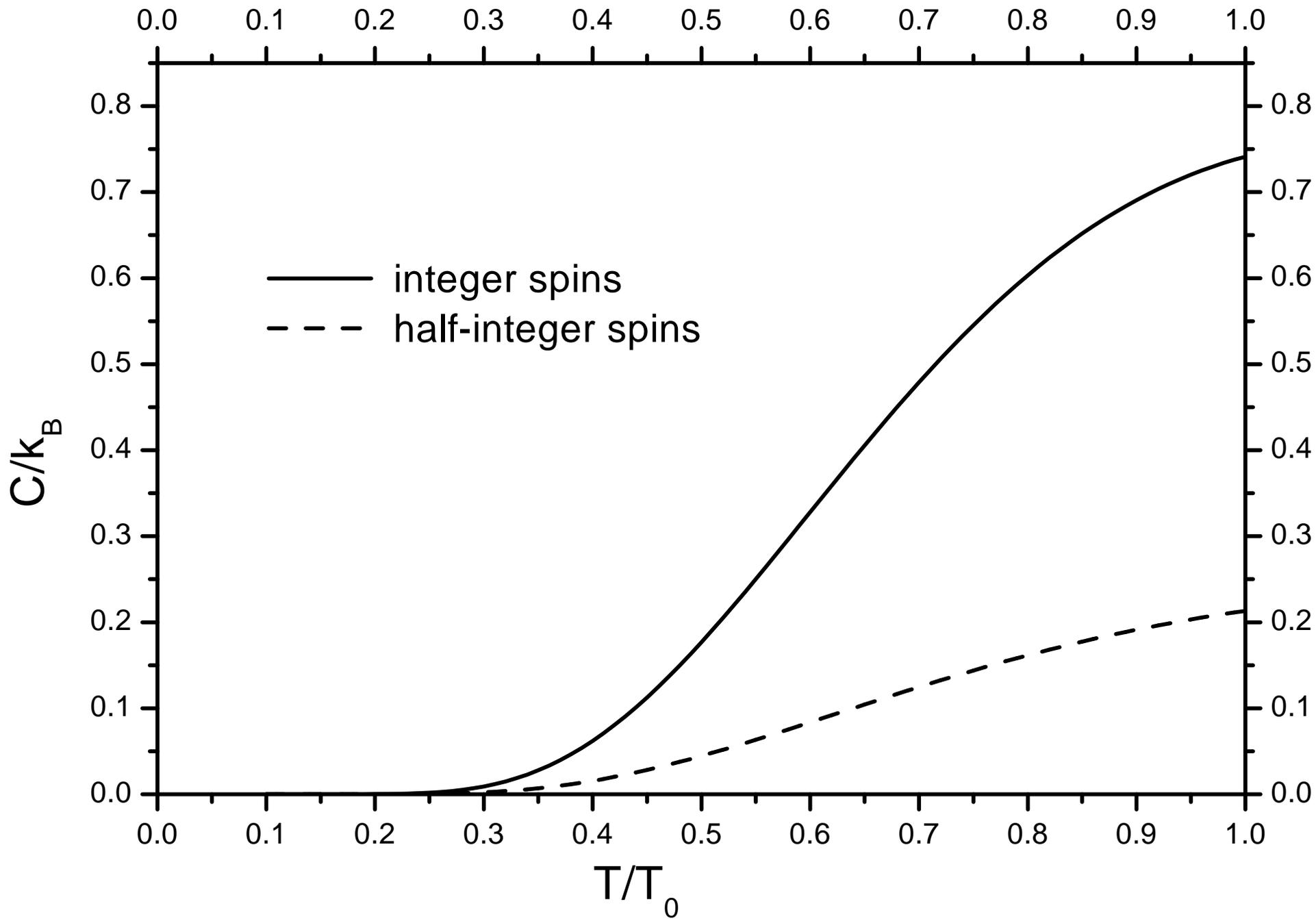
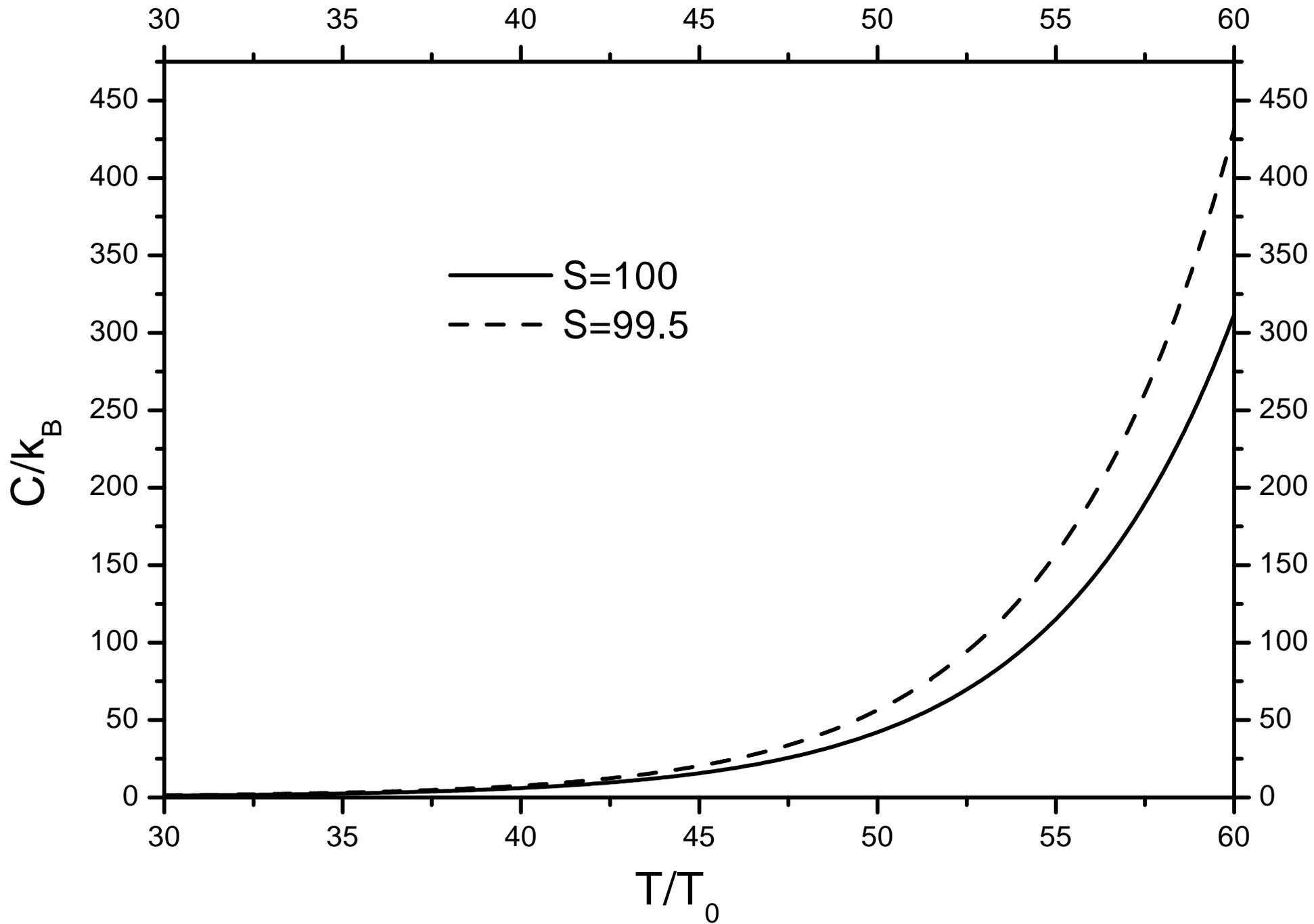
\end{document}